\documentclass[authoryear]{elsarticle}
\usepackage{times} 
\usepackage{natbib}
\usepackage{booktabs,caption,fixltx2e}
\usepackage[utf8]{inputenc}
\usepackage[multiple]{footmisc}
\usepackage[flushleft]{threeparttable}
\usepackage{color}
\usepackage{amsmath}
\usepackage{amsfonts}
\usepackage{hyperref}
\usepackage{setspace}
\makeatletter
\newcommand\footnoteref[1]{\protected@xdef\@thefnmark{\ref{#1}}\@footnotemark}
\usepackage{ragged2e}
\justifying
\title{Mid-infrared vibrational study of deuterium-containing PAH variants}
\author
{
Mridusmita Buragohain$^{1}$\mbox{*}, 
Amit Pathak$^{1}$, 
Peter Sarre$^2$, 
Takashi Onaka$^{3}$ and 
Itsuki Sakon$^{3}$
\vspace*{10pt}\\
$^{1}$Department of Physics, Tezpur University, Tezpur 784\,028, India (ms.mridusmita@gmail.com, amit@tezu.ernet.in, \mbox{*} Corresponding author)\\
$^{2}$School of Chemistry, The University of Nottingham, University Park, Nottingham, NG7~2RD, United Kingdom\\
$^{3}$Department of Astronomy, Graduate School of Science, The University of Tokyo, Tokyo 113-0033, Japan}

\begin{document}
\date{}
\begin{frontmatter}
\begin{abstract}
Polycyclic Aromatic Hydrocarbon (PAH) molecules have been long proposed
to be a major carrier of ‘Unidentified Infrared’ (UIR) emission bands that
have been observed ubiquitously in various astrophysical environments.
These molecules can potentially be an efficient reservoir of deuterium.
Once the infrared properties of the deuterium-containing PAHs are well
understood both experimentally and theoretically, the interstellar
UIR bands can be used as a valuable tool to infer the cause of  the deuterium
depletion in the ISM.

Density Functional Theory (DFT) calculations have been carried out on deuterium-containing ovalene variants to 
study the infrared properties of these molecules. These include deuterated ovalene, cationic deuterated ovalene, 
deuteronated ovalene and deuterated-deuteronated ovalene. We present a D/H ratio calculated 
from our theoretical study to compare with the observationally proposed D/H ratio.
\end{abstract}

\end{frontmatter}

\section{Introduction}

Polycyclic Aromatic Hydrocarbon (PAH) molecules have long been recognized 
as carriers of the mid-infrared emission bands popularly known as Unidentified 
Infrared (UIR) emission bands, which are observed at 3.3, 6.2, 7.7, 8.6, 11.2 
and 12.7 $\mu \rm m$ and longer wavelengths as broad emission features towards diverse 
astrophysical sources \citep[]{Tielens08}. After being discovered by \citet[]{Gillett73} 
in the 8-13 $\mu \rm m$ range, these infrared (IR) bands have been observed towards 
numerous Galactic and extragalactic sources that vary in physical 
and chemical environments \citep[]{Onaka96, Mattila96, Verstraete96, Hony01, 
Verstraete01, Peeters02, Abergel02, Acke04, Sakon04}. Increasing number of IR 
observations indicate that widespread, extremely stable interstellar particles 
are responsible for such features \citep[]{Allamandola89}. The search for the 
origin of these features began with the hypothesis of \citet[]{Duley81} that
vibrations of chemical functional groups attached to small carbon grains might 
produce such features. \citet[]{Sellgren84} proposed transiently heated very small grains (size $\sim$ 10\AA{})
to be the carriers.
Later, \citet[]{Leger84} and \citet[]{Allamandola85}
independently suggested that the features arise due to the vibrational relaxation 
of PAH molecules on absorption of background UV photons, giving rise to infrared 
fluorescence. However, identification of the exact molecular form of PAH has not been successful so far. 
Recent discoveries of some weak features in the $3-20~\mu \rm m$ range with the Short Wavelength Spectrometer (\textit{SWS}) 
on board \textit{ISO} point to the existence of an extended PAH family rather than a single form that may 
explain UIR features \citep[]{Verstraete96, Peeters04}. 

PAH molecules bear a significant fraction of interstellar carbon of about 5-10 \% \citep[]{Tielens08} and take part in crucial 
chemical processes, including heating of the ISM through the photo-electric effect and in determining the charge balance inside molecular 
clouds \citep[]{Lepp88, Verstraete90, Bakes94, Peeters04}. PAH molecules are also proposed to be potential carriers 
of `Diffuse Interstellar Bands' (DIBs) that are absorption features on the interstellar extinction 
curve \citep[]{Crawford85, Leger85, Salama96, Salama11, Cox06, Pathak08}. The recent identification of two DIBs 
with C$_{60}^+$ strongly supports the existence of large-sized PAHs in the 
ISM \citep[]{Campbell15, Ehrenfreund15}. \citet[]{Berne15} recently proposed the top-down formation
of fullerene (C$_{60}$) by dehydrogenation of large PAHs using a photochemical model.
In addition, a recent observation has proposed that some of the interstellar deuterium 
(D) exists in the form of PAD or D$\rm{_n-}$PAH; i.e. a PAH molecule with deuterium attached \citep[]{Peeters04}. PAHs are likely 
to accommodate some of the primordial D \citep[]{Draine06} which may explain the present lower value of D/H 
in the interstellar gas. Due to the potential astrophysical as well as environmental implications, 
extensive study on regular and substituted PAHs has been carried out both experimentally and 
theoretically \citep[]{Langhoff96, Hudgins98a, Hudgins98b, Allamandola99, Oomens03, Mattioda05, Pathak08, Bauschlicher10,
 Galue10, Simon11}.

Current research on substituted PAHs suggests the presence of deuterated PAHs (PADs and D$\rm_n$PAHs) as possible carriers
for some of the observed UIRs \citep[]{Peeters04, Draine06, Onaka14, Mridu15, Doney15}. Such species are crucial as they show characteristic
features in the $4-5~\mu \rm m$ region which is a featureless region for other pure as well as known substituted PAHs.
Along with deuterated PAHs, other forms of deuterium-containing PAH variants are equally important
as they are expected to show similar features as that of deuterated PAH. In our latest 
report, \citep[]{Mridu15}, a new form of deuterium-containing PAH molecule, i.e. deuteronated PAHs (DPAH$^+$) 
have been suggested as potential carriers for observed bands in the $4-5~\mu\rm m$ region. Deuteronated
PAH molecules of increasing size have been discussed in \citet[]{Mridu15}. In the present work, possible 
variants of deuterium-containing ovalene have been studied to determine the 
expected vibrational transitions and to compare with observational data.

\section{Deuterium-containing PAH variants}

A regular PAH molecule can be converted into a deuterated PAH by exchange of D from $\rm{D{_2}O}$ ice with one of the peripheral
hydrogens in PAH in presence of UV radiation \citep[]{Sandford2000}. There are other formation mechanisms as discussed 
by \citet[]{Tielens83, Tielens92, Tielens97} and \citet[]{Allamandola87, Allamandola89}. A deuterated PAH molecule may
be converted into an alternative form of molecule e.g. deuteronated PAH ($\rm{DPAH^{+}}$, a PAH with a deuteron added to it)
in the course of chemical reactions occurring in the ISM, through addition of D to PAH radical cations or low temperature
ion-molecule reaction followed by deuterium fractionation (see \citet[]{Mridu15} for details). 
Spectral evidence of such species in the ISM has been discussed recently by \citet[]{Peeters04, Onaka14, Mridu15}.
\citet[]{Peeters04} reported two bands in the region of $4-5~\mu \rm m$ associated with deuterated PAHs. The bands at
$4.4~\mu \rm m$ and $4.65~\mu \rm m$ have been suggested to arise from the stretching of aromatic and aliphatic $\rm C-D$ bonds respectively 
in a deuterated PAH (PAD or D$\rm_n$PAH). These two bands are analogous to the 3.3 and 3.4~$\mu \rm m$ bands which
are characteristics of aromatic and aliphatic C-H stretching, respectively. The idea has gained further support from an agreement 
of observed D/H ratio \citep[]{Peeters04} with the proposed
value of D/H ratio by \citet[]{Draine06}. According to \citet[]{Draine06}, some of the primordial D atoms are incorporated into PAHs 
which may explain the current lower value of D/H in the interstellar gas
compared with the primitive value of D/H. \citet[]{Draine06} proposed a D/H ratio of $\sim$~0.3 for PAHs 
which agrees with the observations of \citet[]{Peeters04}. \textit{AKARI} 
observations \citep[]{Onaka14} tentatively detected the $\rm{C-D}$ band vibrations at 4.4 and 4.65 $\mu \rm m$ and give an upper limit for the D/H ratio. 
These observations favor a much lower D/H ratio and suggest that D is more likely to be accommodated in large PAHs.

\section{Computational Approach}

To assign carriers for the large set of observed UIR bands, comparison between observational, experimental 
and theoretical spectra of PAHs is of the utmost importance. However, experimental spectroscopy has limitations 
especially for large and complex PAHs for several reasons.Synthesis of large PAHs
and reproducing interstellar conditions in the laboratory are the major constraints faced 
in experimental spectroscopy. Theoretical quantum chemical 
calculations provide the missing link. Density Functional Theory (DFT) is immensely valuable to study 
the vibrational properties of a large set of PAH candidate carriers 
\begin{figure*}[h]
\includegraphics[width=9cm, height=12cm]{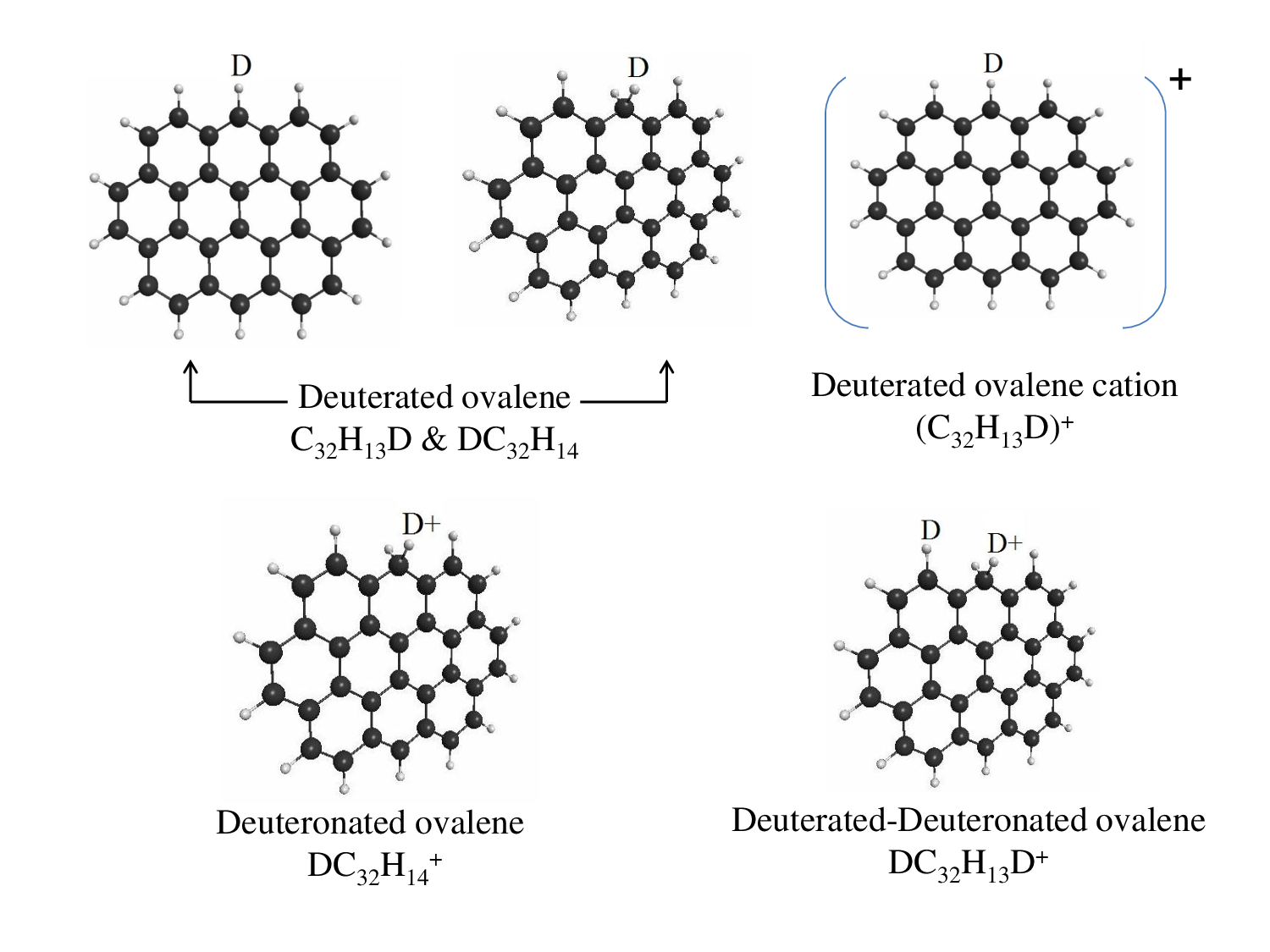}
\caption{examples of deuterium-containing ovalene variants}
\label{fig1}
\end{figure*}
in relation to UIR bands \citep[]{DeFrees93, Langhoff96, Szczepanski96, 
Bauschlicher97, Bausch97b, Langhoff98, Hudgins01, Hudgins04a, 
Pathak05, Pathak06, Pathak07, Malloci07a, Malloci08, Pauzat11, Klaerke13, Candian14, Roser14, Mridu15}. 
In this work, a DFT combination of B3LYP/6-311G** has been used to optimize the molecular 
structure of PAHs. Frequencies and intensities of vibrational transitions have been computed using the optimized geometry
at the same level of calculation. The calculated intensities are used as input 
to an emission model \citep[]{Cook98, Pech02, Pathak08a}. The emission model uses black body radiation
generated by a source having effective temperature of 40,000~K. The PAH molecule absorbs this
radiation following their respective absorption cross-section. The excited PAH
attains a peak temperature and cools down following a cascade mechanism emitting IR photons
corresponding to their vibrational modes. These IR photons are added up to generate the emission
spectrum. Details of the emission model may be found in \citet[]{Pathak08a}. The emitted energy and frequency thus obtained from
emission model is used to plot a Gaussian profile of FWHM 30 cm$^{-1}$. The profile width is typical for PAHs emitting
in interstellar environment and depends on vibrational redistribution of the molecule \citep[]{Allamandola89}. 
Considering the emission model with a lower black body effective temperature of say 30,000~K, 
we find no difference in the PAH emission spectrum except that the emitted energy is slightly reduced.

It is important to note that the theory overestimates the experimental frequency. 
To bring the frequencies into accordance with experimental values, theoretical frequencies are scaled down by using 
three different scaling factors for three different ranges of vibrational frequency. Scaling factors are calculated by 
comparing the frequencies of selected PAH molecules with available experimental data \citep[]{Mridu15}.
Scaling factors used here are 0.974 for $\rm{C-H}$ out of plane (OOP), 0.972 for $\rm{}C-H$ in-plane and $\rm{C-C}$ stretching 
and 0.965 for $\rm{C-H}$ stretching \citep[]{Mridu15}. In case of deuterium-containing PAHs, since we do 
not have laboratory data for aliphatic deuterium-containing PAHs, a scaling factor of 0.965 corresponding
to $\rm{C-H}$ stretching is used for  $\rm{C-D}$ stretching. The shift in frequencies of aliphatic bonds needs experimental support 
and may have some uncertainty. 
Relative intensities (Int$\rm_{rel}$ \footnote{Int$\rm_{rel}$
=$\frac{\rm{absolute~intensity}}{\rm{maximum~absolute~intensity}}$}) are obtained by taking the ratio of all intensities to the maximum intensity near 
3060 cm$^{-1}$ for neutral ovalenes. Similarly, for cations, we divide all the intensities by maximum intensity 
that appears near 1600 cm$^{-1}$ for normalization. Several unresolved bands might add up resulting some band intensity to cross unity. 


A mid-sized PAH ovalene ($\rm C_{32}H_{14}$) is chosen for this work.
Figure~\ref{fig1} shows the structure of various deuterium-containing ovalene variants.
This work includes DFT study of i) Deuterated ovalene (aromatic), ii) Deuterated ovalene (aliphatic), 
iii) Deuterated ovalene cation iv) Deuteronated ovalene, v) Deuterated-Deuteronated ovalene. Isomers of all the sample
molecules are also considered in this work. The data presented here were produced using the 
QChem quantum chemistry suite of programs \citep[]{YSaho15}. The vibrational modes of the molecule
are identified using graphical software available for computational chemistry packages.

\section{Results and Discussion}
\subsection*{Neutral and cationic Ovalene}
 \begin{figure*}[h]
\centering
\includegraphics[width=10cm, height=11cm]{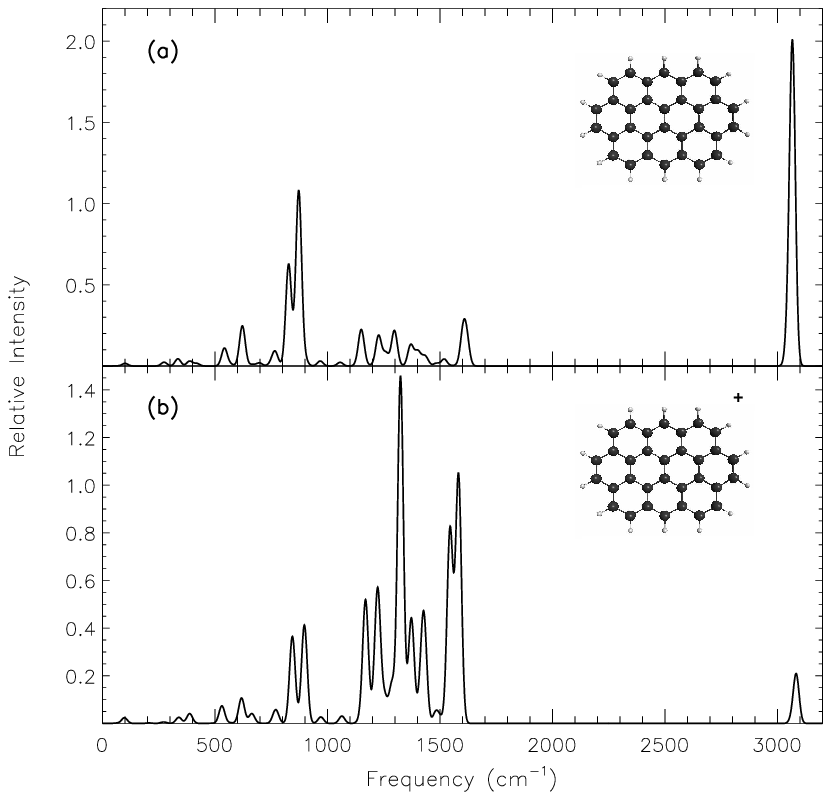}
\caption{Theoretical emission spectra of (a) neutral ovalene (C$_{32}$H$_{14}$), 
(b) ovalene cation ($\rm C_{32}H_{14}^+$)}
\label{fig2}
\end{figure*}
These are pure ovalenes in neutral and ionized forms. The theoretically computed emission spectra
of neutral ovalene (C$_{32}$H$_{14}$) and cationic ovalene ($\rm C_{32}H_{14}^+$) are presented in Figure~\ref{fig2}.
Neutral ovalene shows strong lines at $\sim~800-900~\rm cm^{-1} (\sim~13-11~\mu \rm m$) and 
$\sim3050~\rm cm^{-1} (\sim~3.3~\mu \rm m$) due to $\rm C-H$ oop and $\rm C-H$ stretching vibrational modes, respectively (Figure~\ref{fig2}a).
Weak features at $\sim1000-1600~\rm cm^{-1} (\sim10-6~\mu \rm m$) are characteristics of $\rm C-H$ in-plane 
bending and $\rm C-C$ stretching vibrational modes. These features are inherent for any form of neutral PAH molecules.
Likewise, any form of cationic PAH molecule shows a greater number of features in the $\sim~1000-1600~\rm cm^{-1} 
(\sim~10-6~\mu \rm m$) region as shown in Figure~\ref{fig2}b. The region below $\sim~1000~\rm cm^{-1}$ is attributed to $\rm C-H$ oop modes.
Unlike neutral PAHs, all cationic PAHs show a weak feature at $3080~\rm cm^{-1},\sim~3.3~\mu \rm m$ 
due to $\rm C-H$ stretching vibrational modes. These are the standard features expected from any neutral and ionized PAH 
molecule.\footnote{For converting $\rm cm^{-1}$ to $\mu \rm m$, $\frac{10,000}{\rm cm^{-1}}$ is a simple conversion formula.}
\subsection*{Deuterated Ovalene}
\begin{figure*}
\centering
\includegraphics[width=12cm, height=16cm]{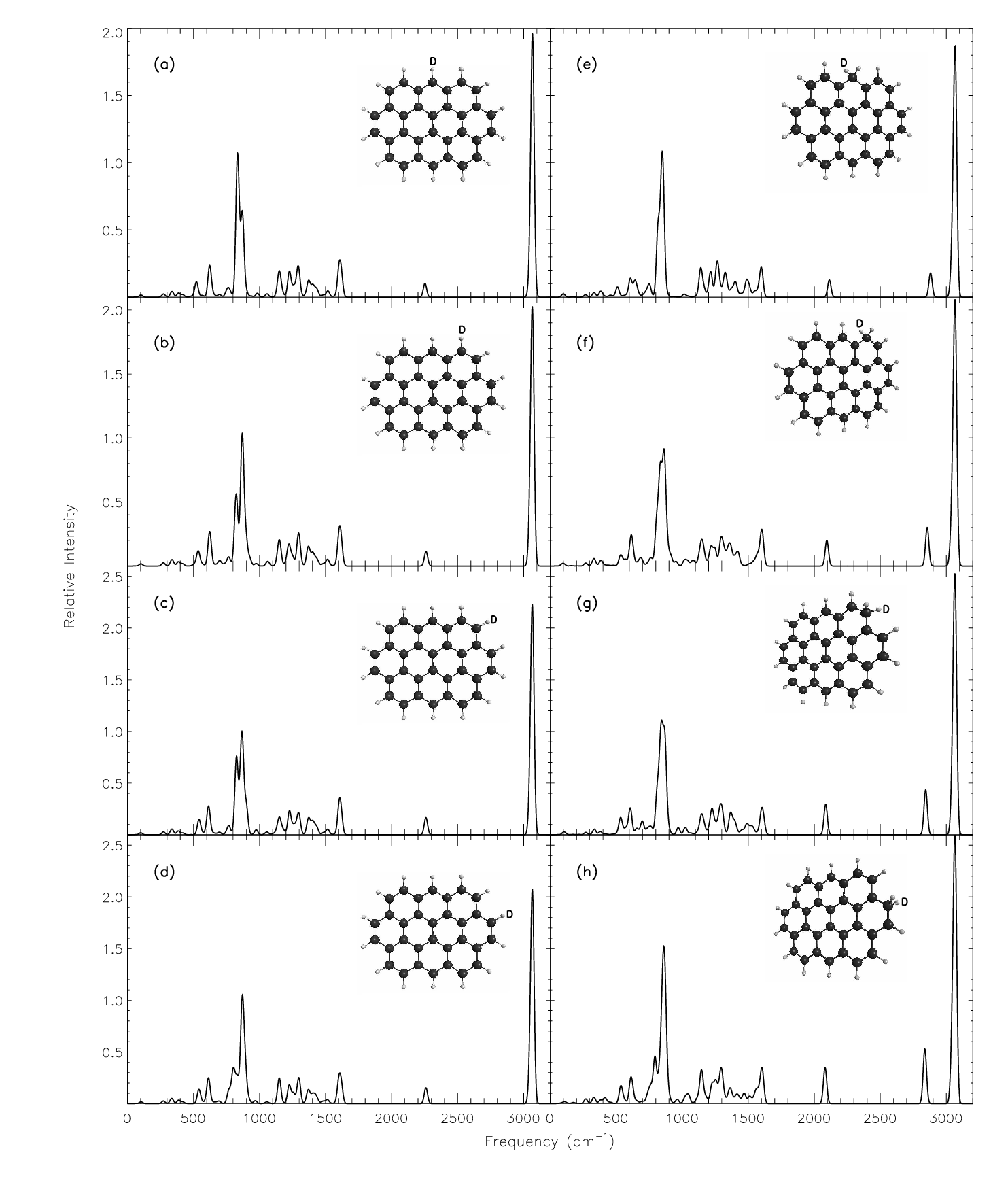}
\caption{Theoretical spectra of (a) deuterated ovalene (aromatic,C$_{32}$H$_{13}$D), (b) isomer 1 of 
C$_{32}$H$_{13}$D, (c) isomer 2 of C$_{32}$H$_{13}$D, (d) isomer 3 of C$_{32}$H$_{13}$D, 
(e) deuterated ovalene (aliphatic, DC$_{32}$H$_{14}$), (f) isomer 1 of DC$_{32}$H$_{14}$,
(g) isomer 2 of DC$_{32}$H$_{14}$, (h) isomer 3 of DC$_{32}$H$_{14}$}
\label{fig3}
\end{figure*}

A deuterium atom can attach to the periphery of a neutral PAH molecule either in an aromatic or aliphatic site
that leads to the formation of a deuterated PAH molecule (PAD or D$\rm_{n}$PAH) \citep[]{Peeters04}. Figure~\ref{fig3} 
shows the emission spectra of deuterated ovalene, both aromatic \footnote{D replacing H atom so that the 
resulting $\rm C-D$ bond remains aromatic in nature} and aliphatic \footnote{addition of an extra D to neutral ovalene so that the 
resulting $\rm C-D$ bond remains aliphatic in nature} (C$_{32}$H$_{13}$D and 
DC$_{32}$H$_{14}$). Ovalene has four unique sites of deuteration which give four isomers 
of deuterated ovalene, both aromatic and aliphatic. In Figure~\ref{fig3}, the emission spectra 
of all the isomers of deuterated ovalene are also presented.

Both aromatic and aliphatic deuterated ovalene along with its isomers show usual features similar to those present in neutral 
ovalene. Substitution of D reduces the symmetry of 
ovalene that activates modes that were IR-inactive in neutral ovalene \citep[]{Mulas03}.
In deuterated ovalenes, weak $\rm C-D$ oop ($\sim600-700~\rm cm^{-1},\sim16-14~\mu \rm m$) and $\rm C-D$ in-plane modes
($\sim 860-900~\rm cm^{-1},\sim~11~\mu \rm m$) appear due
to the presence of D in the structure. These modes 
are red shifted compared to analogous $\rm{C-H}$ modes due to the heavier deuterium. 
Figure~\ref{fig3} (a-d) are representatives of the same molecule, i.e. aromatic deuterated ovalene, but differ only in the position of D.
In Figure~\ref{fig3} (a-d), a unique feature appears at $\sim~2260~\rm cm^{-1}$, 4.4$~\mu \rm m$ \footnote{wavenumbers are averaged
among the isomers of the same molecule} 
(Int$\rm_{rel(em)}$~0.13 \footnote{Relative intensity from emission model, average is taken for the same molecule including its isomers}) 
which is characteristic of $\rm C-D$ stretching in C$_{32}$H$_{13}$D. 
This feature is analogous to the strong $\rm C-H$ stretching modes at $\sim~3.3~\mu \rm m$. 
In the case of DC$_{32}$H$_{14}$ and its isomers, as shown in Figure~\ref{fig3} (e-h), addition of an extra deuterium 
atom does not affect the spectrum much except in the $4-5~\mu \rm m$ region.
Addition of deuterium to a neutral ovalene molecule breaks the aromatic nature at the site where D
is added. Stretching of the $\rm{C-D}$ and $\rm{C-H}$ aliphatic bonds in DC$_{32}$H$_{14}$ gives features at
$\sim~2095~\rm cm^{-1}, 4.8~\mu \rm m$ (Int$\rm_{rel(em)}~0.24$) and $2880.31~\rm cm^{-1}, 
3.5~\mu \rm m$ (Int$\rm_{rel(em)}~0.36$) respectively. Table~\ref{tab1} lists intensities and positions of 
$\rm C-D$ stretching transitions for all the deuterated ovalenes (both aromatic and aliphatic) along with its isomers. The positions and
intensities of $4.4~\mu \rm m$ (aromatic $\rm C-D$ stretching) and $4.7/4.8~\mu \rm m$ (aliphatic $\rm C-D$ stretching) 
feature are partially affected by the position of D. There is
a maximum wavenumber variation of 8 $\rm cm^{-1}$ for aromatic deuterated ovalene and 32 $\rm cm^{-1}$ for aliphatic deuterated
ovalene. Similarly, Int$\rm_{rel(em)}$ also varies among the isomers, however no uniform pattern is observed (Figure~\ref{fig3}).

\subsection*{Deuterated Ovalene cation and Deuteronated Ovalene}
\begin{figure*}
\includegraphics[width=13cm, height=17cm]{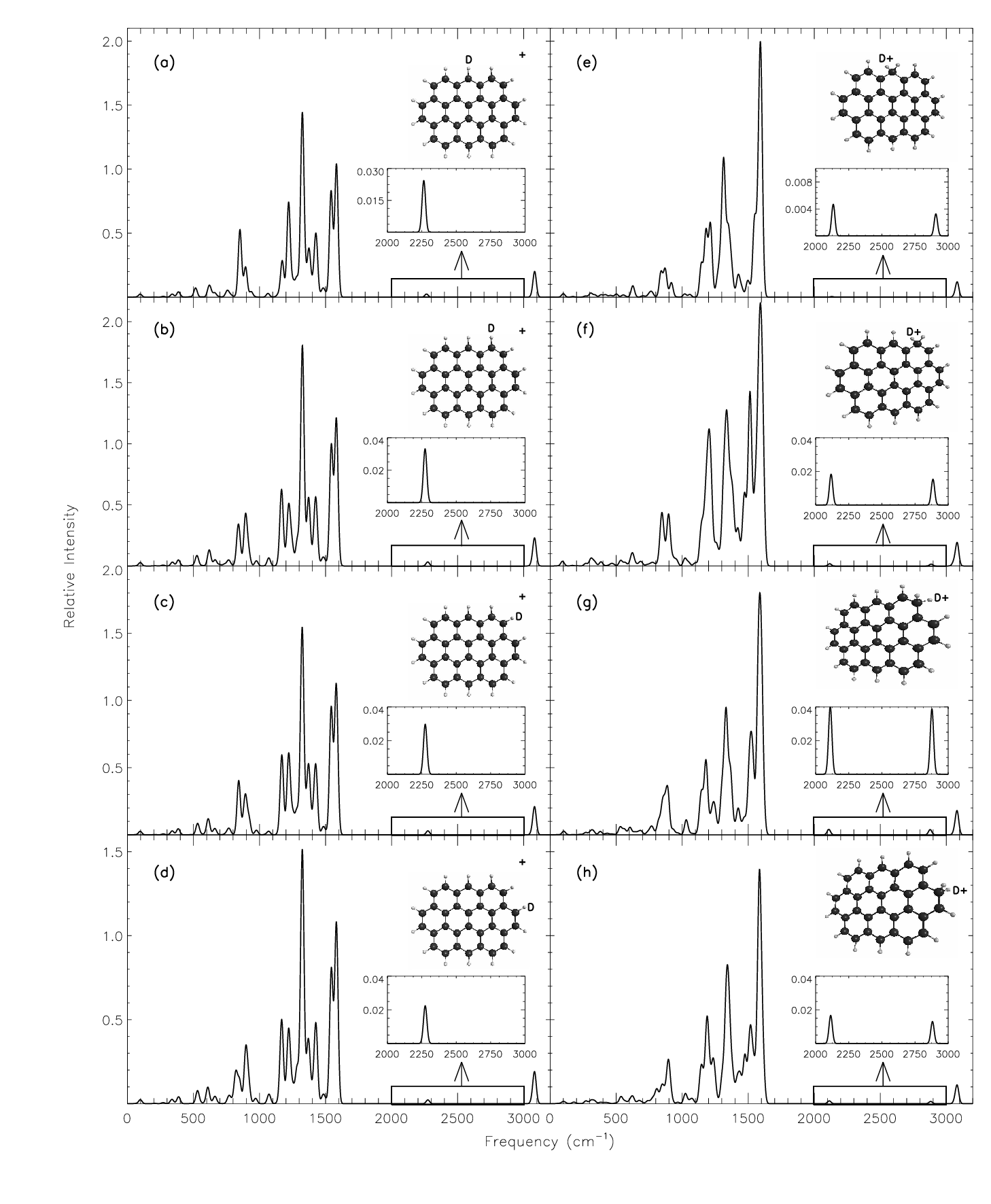}
\caption{Theoretical spectra of (a) deuterated ovalene cation, (C$_{32}$H$_{13}$D$^+$), (b) isomer 1 of 
C$_{32}$H$_{13}$D$^+$, (c) isomer 2 of C$_{32}$H$_{13}$D$^+$, (d) isomer 3 of C$_{32}$H$_{13}$D$^+$, 
(e) deuteronated ovalene, (DC$_{32}$H$_{14}^+$), (f) isomer 1 of DC$_{32}$H$_{14}^+$,
(g) isomer 2 of DC$_{32}$H$_{14}+$, (h) isomer 3 of DC$_{32}$H$_{14}^+$}
\label{fig4}
\end{figure*}
All the isomers of cationic forms of deuterated ovalene ($\rm{C_{32}H_{13}D^+}$) show more features compared to their 
neutral counterparts; particularly in the region $\sim~1000-1600~\rm cm^{-1},\sim~10-6~\mu \rm m$ as shown in Figure~\ref{fig4} (a-d). 
This region is characteristics of $\rm C-H$ in-plane and $\rm C-C$ stretching vibrational modes. This is similar to that of ovalene
cation, $\rm{C_{32}H_{14}^+}$ (Figure~\ref{fig2}b). The effect of deuterium is however not observed in this region. 
The region below $\sim~1000~\rm cm^{-1}$ is attributed to $\rm C-H$ oop modes, 
which is again free from any significant D-associated modes. In Figure~\ref{fig4} (a-d),
a weak feature at $\sim~2272~\rm cm^{-1}, 4.4~\mu \rm m$ (Int$\rm_{rel(em)}$ 0.03) is 
attributed to the aromatic $\rm C-D$ stretching mode in $\rm C_{32}H_{13}D^+$. The $2000-3000~\rm cm^{-1}$
($5-3\mu \rm m$) region is magnified to highlight the weak features. Unlike the neutral counterparts of deuterated ovalene, 
cationic deuterated ovalenes show a weak feature at $3080~\rm cm^{-1},\sim~3.3~\mu \rm m$. The position of the $4.4~\mu \rm m$ 
band is not much affected by the position of D in solo and duo site with a maximum wavenumber separation of 
$11~\rm cm^{-1}$. Similarly, Int$\rm_{rel(em)}$ does not vary much as can be seen in Table~\ref{tab1}.

A new form of PAH candidate carrier of mid infrared emission bands; deuteronated PAH (DPAH$^+$) has recently been
discussed by \citet[]{Mridu15}. Structurally, a deuteronated PAH is a PAH with a deuteron added to 
its periphery. Formation of these PAHs is favorable in the ionized ISM. Its closed shell electronic 
structure makes a deuteronated PAH molecule chemically less reactive than the corresponding PAH cation which is an open shell structure. However, photostability does not change much between similar open shell and closed shell species.  Theoretical IR spectra of 
deuteronated ovalene (DC$_{32}$H$_{14}^+$) and its isomers are presented
in Figure~\ref{fig4} (e-h). Addition of a deuteron (D$^+$) to a neutral PAH at different positions reduces the symmetry and 
a rich IR spectrum, particularly in the region $\sim~1000-1600~\rm cm^{-1},\sim~10-6~\mu \rm m$ is noted. A number of features
in this region are contributed by $\rm C-H$ in-plane and $\rm C-C$ stretching modes. However, a new D-associated feature
is observed at $1212.41~\rm cm^{-1},~8.3~\mu \rm m$ (Int$\rm_{rel(em)}~0.41$), which
arises due to a combination of $\rm D-C-H$ oop, $\rm C-H$ in-plane and $\rm C-C$ stretching vibrations. Prominent 
features with Int$\rm_{rel(em)}$ between 0.1 and 0.2 due to $\rm C-H$ oop modes are present in the
$\sim~800-900~\rm cm^{-1},\sim~13-11~\mu \rm m$ region. This region is not affected by any significant D-associated modes.
A pure D-associated mode appears at $\sim~2118~\rm cm^{-1},$ i.e., 4.7~$\mu \rm m$ with a very low Int$\rm_{rel(em)}$ of 0.02. This 
feature is attributed to the stretching of the aliphatic $\rm C-D$ bond in DC$_{32}$H$_{14}^+$.
Stretching of aliphatic and aromatic $\rm C-H$ bonds gives weak features at $\sim~2887~\rm cm^{-1} (3.5~\mu \rm m)$ and 
$\sim~3080~\rm cm^{-1} (3.3\mu \rm m)$, respectively. In Figure~\ref{fig4}, the $2000-3000~\rm cm^{-1}$ region is zoomed in to show 
the weak features. Isomers of DC$_{32}$H$_{14}^+$ show a variation in the position of the 4.7~$\mu \rm m$
with a maximum wavenumber variation of $23~\rm cm^{-1}$. Intensities remain almost the same irrespective of the structural difference of
DC$_{32}$H$_{14}^+$. 
\subsection*{DovaleneD$^+$}
DovaleneD$^+$ (DC$_{32}$H$_{13}$D$^+$) carries two types of $\rm C-D$ bonds, aromatic at the 
addition site of D and aliphatic at the addition site of D$^+$. The computed IR spectra of 
\begin{figure*}[h]
\centering
\vspace*{-10cm}
\includegraphics[width=17cm, height=20cm]{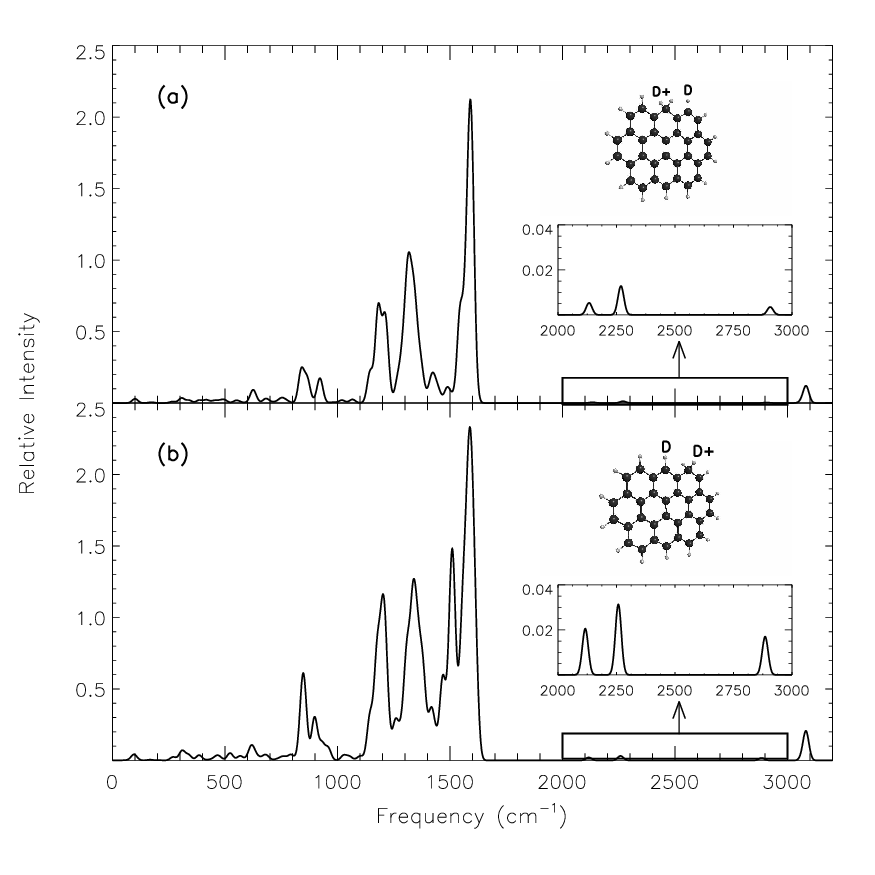}
\caption{Theoretical spectra of DovaleneD$^+$ (DC$_{32}$H$_{13}$D$^+$) and its isomer}
\label{fig5}
\end{figure*}
of DovaleneD$^+$ and its isomer are shown in Figure~\ref{fig5} (a,b). Several isomers 
for DovaleneD$^+$ are possible, though only two are randomly chosen. The structure of DovaleneD$^+$ 
is a combination of deuterated ovalene (aromatic) and deuteronated ovalene 
and shows similar characteristic vibrational modes that are present in deuterated
ovalene (aromatic) and deuteronated ovalene. The spectrum (Figure~\ref{fig5}) is dominated by 
rich $\rm C-H$ in-plane and $\rm C-C$ stretching modes that appear in the 
$\sim1000-1600~\rm cm^{-1},\sim10-6~\mu \rm m$ region. $\rm C-H$ oop modes are 
comparatively weak and are distributed in the $\sim~600-900~\rm cm^{-1},
\sim~16-11~\mu \rm m$ region. The presence of deuteriums in DovaleneD$^+$ causes 
new features to appear in the $\sim~600-1600~\rm cm^{-1},\sim~16-6~\mu \rm m$ region.
The most significant features (above relative intensity 0.05) are $\rm C-D$ 
in-plane and $\rm D-C-H$ oop. The former is analogous to the $\rm C-H$ in-plane mode, 
but is redshifted due to the heavier mass of D and blended with the $\rm C-H$ oop modes. 
This feature appears at $\sim~860~\rm cm^{-1} (11.6~\mu \rm m$, Int$\rm_{rel(em)}~0.08)$ 
in DovaleneD$^+$ and at $\sim~850~\rm cm^{-1} (11.8~\mu \rm m$, Int$\rm_{rel(em)}~0.15$) in its isomer. 
The latter ($\rm{D-C-H}$ oop) appears at $\sim~1212~\rm cm^{-1}$ (8.3~$\mu \rm m$, Int$\rm_{rel}$ 0.42) in DovaleneD$^+$
and at $\sim1200~\rm cm^{-1}$ (8.3~$\mu \rm m$, Int$\rm_{rel}$ 0.66) in its isomer. These features are
however not pure and are mixed with the $\rm C-H$ in-plane and the $\rm C-C$ stretching modes. 
Pure contributions of D atoms in DovaleneD$^+$ are observed at $\sim~2133~\rm cm^{-1} 
(4.7~\mu \rm m$, Int$\rm_{rel}$ 0.005) and at $\sim~2269~\rm cm^{-1} (4.4~\mu \rm m$, Int$\rm_{rel}$ 0.01).
\begin{figure*}
\centering
\vspace*{-10cm}
\includegraphics[width=17cm, height=22cm]{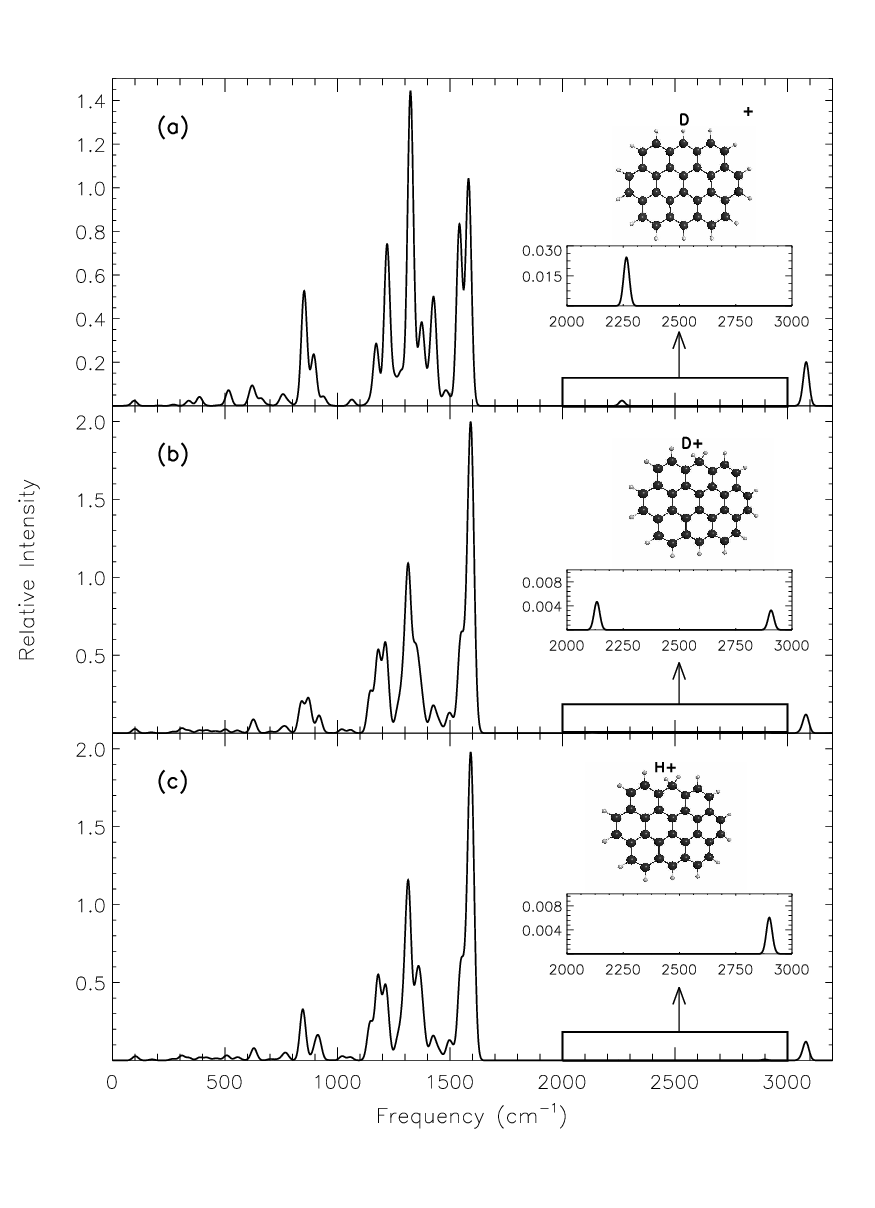}
\caption{Theoretical spectra of (a) deuterated ovalene cation, (C$_{32}$H$_{13}$D$^+$), (b) deuteronated ovalene, (DC$_{32}$H$_{14}^+$)
and protonated ovalene (HC$_{32}$H$_{14}^+$)}
\label{fig6}
\end{figure*}
These two features arise due to the stretching of aliphatic $\rm C-D$ and aromatic $\rm C-D$ bonds 
respectively and are extremely weak compared to $\rm{C-C}$ stretching modes. The stretching of the aliphatic
$\rm C-H$ bond gives rise to a very weak feature at $\sim~2907~\rm cm^{-1} (3.4~\mu \rm m$, 
Int$\rm_{rel}$ 0.004). In the isomer counterpart (Figure~\ref{fig5}b), stretching of aliphatic $\rm C-D$, aromatic $\rm C-D$
and aliphatic $\rm C-H$ bonds are seen at $\sim~2117~\rm cm^{-1} (4.7~\mu \rm m$), $\sim~2258~\rm cm^{-1} (4.4~\mu \rm m$)
and $\sim~2885~\rm cm^{-1} (3.5~\mu \rm m$) with Int$\rm_{rel}$ 0.02, 0.03 and 0.02 respectively.
The region of $2000-3000~\rm cm^{-1}$ is zoomed-in
as shown in Figure~\ref{fig5}, to highlight the weak features at $4.7, 4.4~\&~3.5 ~\mu \rm m$.
As expected, another weak but distinct feature appears at $\sim~3080~\rm cm^{-1} 
(3.3~\mu \rm m$) which is due to aromatic $\rm C-H$ stretching vibrational modes and is inherent in all cationic PAHs.

DovaleneD$^+$ has a possibility of several isomers and a slight variation is present in the
the aliphatic and aromatic C$-$D stretching bands depending on the position of substitution/addition of the 
D atom. This leads to the broadening of the C$-$D stretching band if contribution from all the isomers is considered.

Figure~\ref{fig6} compares the spectra of deuterated ovalene cation and deuteronated ovalene with that of protonated ovalene
(HC$_{32}$H$_{14}^+$). A protonated PAH is a PAH with a proton added to 
its periphery and is structurally identical to a deuteronated PAH. We have considered protonation only at one position and 
compared with respective counterparts of deuterated ovalene cation and deuteronated ovalene. The emission spectrum
of protonated ovalene shows similar features that are present in deuteronated ovalene. An exception is the appearance of a weak
feature at $\sim~2900~\rm cm^{-1} (3.5~\mu \rm m$, Int$\rm_{rel}$ 0.006) due to symmetric H-C-H stretching at the addition
site. Its associated antisymmetric H-C-H stretching appears at $\sim~2915~\rm cm^{-1} (3.4~\mu \rm m$) with very low relative intensity.
\begin{table*}
 
 \small
 \centering
  \begin{minipage}{120mm}
 \caption{Intensities and positions of the $\rm{C-D}$ stretching mode in deuterium-containing PAH variants}
 \label{tab1}

 \begin{tabular}[c]{c|c|c|c|c|c}
 \hline
 \hline
 PAH &  Frequency & Wavelength & Absolute & Int$\rm_{rel(abs)}$ \footnote{Relative intensity (Int$\rm_{rel}$) from absorption} & 
 Int$\rm_{rel(em)}$ \footnote{Relative intensity (Int$\rm_{rel}$) from emission model}\\
   &   ($\rm cm^{-1}$) & ($\mu \rm m$) & intensity (km/mole) &  &\\ \hline
 Deuterated ovalene \footnote{\label{a}C-D bond is aromatic in nature} & 2253 & 4.44 & 9.816 & 0.08 & 0.10 \\ \hline
 Deuterated ovalene \footnoteref{a} & &  &  & \\
 (isomer 1)   & 2261 & 4.42 &  10.118 & 0.08 & 0.11 \\ \hline
 Deuterated ovalene\footnoteref{a} &  &  &  &  &  \\
 (isomer 2) & 2260 & 4.42 & 13.424 & 0.12 & 0.17  \\ \hline
 Deuterated ovalene\footnoteref{a} &  &  &  &  &  \\ 
 (isomer 3) & 2260 & 4.42 & 13.297 & 0.11 & 0.15 \\ \hline
 Deuterated ovalene \footnote{\label{b}C-D bond is aliphatic in nature} & 2114 & 4.73 & 11.681 & 0.09 & 0.13 \\ \hline
 Deuterated ovalene \footnoteref{b} &  &  &  &  & \\
 (isomer 1) & 2095 & 4.77 & 15.748 & 0.14 & 0.2  \\ \hline
 Deuterated ovalene \footnoteref{b} &  &  &  &  & \\
 (isomer 2) & 2086 & 4.79 & 19.360 & 0.2 & 0.29  \\ \hline
 Deuterated ovalene \footnoteref{b} &  &  &  &  & \\
 (isomer 3) & 2082 & 4.80 & 21.566 & 0.24 & 0.35  \\ \hline
 Deuterated ovalene\footnoteref{a} &  &  &  &  &  \\
 cation & 2264 & 4.42 & 5.166 & 0.03 & 0.02  \\ \hline
 Deuterated ovalene\footnoteref{a} &  &  &  &  &  \\
 cation (isomer 1)& 2274 & 4.40 & 5.887 & 0.04 & 0.03  \\ \hline
 Deuterated ovalene\footnoteref{a} &  &  &  &  &  \\
 cation (isomer 2)& 2275 & 4.40 & 5.641 & 0.03 & 0.03  \\ \hline
 Deuterated ovalene\footnoteref{a} &  &  &  &  &  \\
 cation (isomer 3)& 2274 & 4.40 & 4.782 & 0.02 & 0.02  \\ \hline
 Deuteronated ovalene \footnoteref{b} & 2133 & 4.69 & 1.434 & 0.005 & 0.005 \\ \hline
 Deuteronated ovalene \footnoteref{b} &  & &  &  &  \\
 (isomer 1) & 2117 & 4.72 & 3.561 & 0.02 & 0.02  \\ \hline
 Deuteronated ovalene \footnoteref{b} &  & &  &  &  \\
 (isomer 2) & 2110 & 4.74 & 8.755 & 0.04 & 0.04  \\ \hline
  Deuteronated ovalene \footnoteref{b} &  & &  &  &  \\
  (isomer 3) & 2114 & 4.73 & 5.699 & 0.02 & 0.02  \\ \hline
   & 2269 & 4.41 & 3.616 & 0.01 & 0.01 \\
  DovaleneD$^+$ \footnoteref{a}$^{,}$\footnoteref{b} & &  &  &  &  \\ 
  & 2133 & 4.69 & 1.460 & 0.006 & 0.005 \\ \hline
  & 2258 & 4.43 & 5.673 & 0.04 & 0.03 \\ 
 DovaleneD$^+$ \footnoteref{a}$^{,}$\footnoteref{b} (isomer) & & & & \\
     & 2117 & 4.73 & 3.609 & 0.02 & 0.02 \\ \hline
 \end{tabular}
 \begin{tablenotes}
 \small
 \item For structures of various isomers of deuterium containing PAHs, please refer to Figure~\ref{fig3}, Figure~\ref{fig4} and Figure~\ref{fig5} \\
 Int$\rm_{rel(abs)}$ is directly calculated from absorption data obtained from DFT calculation and Int$\rm_{rel(em)}$ 
is calculated from the emission model. Int$\rm_{rel(em)}$ shows an average increase of $\sim$~35~\% from Int$\rm_{rel(abs)}$.
    \end{tablenotes}
 \end{minipage}
 
  \end{table*}
 
\section{Astrophysical Implications}
PAHs with incorporated deuterium might be crucial as it might provide an explanation for
the missing primordial D which could not be solely answered by astration\footnote{conversion 
of D into other heavy elements due to nuclear fusion in stellar interiors}.
The current value of D/H has been estimated to be $\sim$~7ppm to $\sim$~22 ppm \citep[]{Jenkins99, Sonneborn2000,
Wood04, Linsky06} along various lines of sight, whereas the primordial D/H ratio is suggested 
as $\sim$~26 ppm \citep[]{Moos02, Steigman03, Wood04}. \citet[]{Draine06}
has proposed that the problem of reduced D/H ratio can be explained if some of the primordial Ds are considered
to be depleted in interstellar dust. Among all forms of interstellar dusts, some might be depleted onto PAHs which 
may produce a deuterated PAH molecule. \citet[]{Draine06} also proposed a D/H ratio of $\sim$~0.3 
in PAHs which is in accordance with the present estimated ratio of D/H in interstellar gas. For observational search
of such deuterium or other deuterium-containing PAH variants in the ISM, spectral observations of these molecules are desired.
 
In this report, we have considered deuterium-containing PAH variants for theoretical
spectroscopic study. All frequencies corresponding to D-associated modes in  
deuterium-containing PAHs may not be used to compare with observations due to 
their low intensity. Another important fact is that $\rm C-D$ in-plane and 
oop modes merge with other usual modes that are present in a pure PAH. Such features are inappropriate 
to distinguish any interstellar deuterium-containing PAH. An exception is the $\rm C-D$ stretching mode 
($4-5~\mu \rm m$) that uniquely may help to identify a PAH candidate with deuterium.
The previous section emphasizes that all
kinds of deuterium-containing PAH variants show unique features in the $4-5~\mu \rm m$ region. 
This region has not been identified with any significant lines apart from the lines 
($4.4~\mu \rm m$ and $4.7~\mu \rm m$) expected from deuterium-containing PAHs
as seen in the emission spectra of PAHs. However, there is a possibility of overtones and combination bands 
occurring at a similar position \citep[]{Allamandola89}. Stretching of aromatic and aliphatic $\rm C-D$ bonds in a deuterium-containing PAH
causes two features to appear at $\sim~4.4~\mu \rm m$ and $\sim~4.7~\mu \rm m$ respectively. Table~\ref{tab1}
describes the position and intensities of the corresponding $\rm{C-D}$ stretching modes in our sample molecules.
The intensity of these features depends on the percentage of deuteration and also on the position of D in the isomers. 
The same features ($4-5~\mu \rm m$ region) have been observationally detected by \citet[]{Peeters04}
at $\sim~4.4~\mu \rm m$ and $\sim~4.65~\mu \rm m$ towards the Orion bar and M17. 
The observed features are assigned to $\rm C-D$ stretching vibrational modes in PADs or D$\rm_n$PAHs. \textit{AKARI}
also observed a small area of overlapping region as that of \citet[]{Peeters04}, and only detected some excess 
emissions at these wavelengths \citep[]{Onaka14}. Bands at $\sim~4.4~\mu \rm m$ and $\sim~4.65~\mu \rm m$
are analogous to bands at $\sim~3.3~\mu \rm m$ (aromatic $\rm C-H$ stretching) and $\sim~3.5~\mu \rm m$ (aliphatic 
$\rm C-H$ stretching) respectively. The $4-5~\mu \rm m$ region is pure in D characteristic 
vibrational modes and may be used to determine the D/H ratio. 
Both the observations made by \textit{ISO} and \textit{AKARI} \citep[]{Peeters04, Onaka14} 
estimated a D/H ratio by taking the ratio of band intensities at 
$\sim~4-5~\mu \rm m$ to $\sim~3-4~\mu \rm m$. \citet[]{Peeters04} proposed D/H ratios of 
0.17$\pm$0.03 in the Orion bar and 0.36$\pm$0.08 in M17.
\citet[]{Onaka14} suggested a comparatively low value of D/H (0.03)  
and proposed low deuteration limited to large PAHs. It is implicitly assumed that a 
molecule can have more than one D \citep[]{Peeters04, Onaka14}.  Thus, 
emission per $\rm{C-D}$ bond was considered in observations \citep[]{Peeters04, Onaka14}. 
This work proposes a [D/H]$\rm_{int}$ that is calculated by taking the ratio of integrated band intensities 
due to $\rm C-D$ stretching to that of $\rm C-H$ stretching from the emission model. Deuterated
ovalene, deuterated ovalene cation and deuteronated ovalene carry only one D atom, whereas
\begin{table*}[h]
 \centering
  \begin{minipage}{120mm}
 \caption{Theoretically computed D/H ratios in deuterium-containing ovalene variants}
 \label{tab2}
 \begin{tabular}[c]{c|c|c}
 \hline \hline
 PAHs &  no of 
 & [D/H]$\rm_{int}$ \footnote[5]{[D/H]$\rm_{int}=$intensity of $\rm{C-D}$ stretch/intensity of $\rm{C-H}$ stretch from the emission model} 
  \\
 &  D atoms, n &   \\ \hline
Deuterated ovalene (aromatic) &  1 & 0.05 \\ \hline
Deuterated ovalene isomer 1 (aromatic) &  1 & 0.07  \\ \hline
Deuterated ovalene isomer 2 (aromatic) &  1 & 0.05  \\ \hline
Deuterated ovalene isomer 3 (aromatic) & 1 & 0.07  \\ \hline
Deuterated ovalene (aliphatic) &  1 & 0.06  \\ \hline
Deuterated ovalene isomer 1 (aliphatic) & 1 & 0.07 \\ \hline
Deuterated ovalene isomer 2 (aliphatic)&  1 & 0.09  \\ \hline
Deuterated ovalene isomer 3 (aliphatic)&  1 & 0.1 \\ \hline
Deuterated ovalene cation &   1 & 0.11   \\ \hline
Deuterated ovalene cation &   1 & 0.13   \\ 
isomer 1  & &  \\ \hline
Deuterated ovalene cation &   1 & 0.13  \\ 
isomer 2 & &   \\ \hline
Deuterated ovalene cation &  1 & 0.12   \\ 
isomer 3  & &  \\ \hline
Deuteronated ovalene &   1 & 0.04  \\ \hline
Deuteronated ovalene isomer 1&   1 & 0.08  \\ \hline
Deuteronated ovalene isomer 2&   1 & 0.17 \\ \hline
Deuteronated ovalene isomer 3&   1 & 0.12  \\ \hline
DovaleneD$^+$ & 2 & 0.14 (0.07)\footnote[6]{\label{aaa}[D/H]$\rm_{abs}$=$\frac{\rm [D/H]_{int}}{\rm n}$}  \\ \hline
DovaleneD$^+$ isomer & 2 & 0.22 (0.11) \footnoteref{aaa} \\ \hline

\end{tabular}
\begin{tablenotes}
 \small
 \item {[D/H]$\rm_{abs}$ is equal to [D/H]$\rm_{int}$ for all molecules except for DovaleneD$^+$ and its isomer as the no of D atom
 is one in deuterated ovalene, deuterated ovalene cation and deuteronated ovalene. DovaleneD$^+$ and its isomer consist of two D atoms
 and thereby give a different [D/H]$\rm_{abs}$ from [D/H]$\rm_{int}$.}
    \end{tablenotes}
\end{minipage}
 \end{table*}
DovaleneD$^+$ has two D atoms. To make a comparative study, 
a [D/H]$\rm_{abs}$ is computed which is nothing
but [D/H]$\rm_{int}$ (intensity of C-D stretch/intensity of C-H stretch) per 
number of D atoms. Table~\ref{tab2} describes the theoretical [D/H]$\rm_{abs}$
ratios obtained from deuterium-containing ovalene variants. Since intensity 
is sensitive to the position in which D is substituted and/or added, [D/H]$\rm_{abs}$ 
shows variation from one isomer to another isomer in all considered PAHs. Such variation is 
salient for deuteronated ovalene. The calculated [D/H]$\rm_{abs}$ values
for deuterated ovalenes are small compared to \textit{ISO} observation and close to \textit{AKARI}
observation. On the contrary, for deuterated ovalene cations, [D/H]$\rm_{abs}$ values are in close proximity to \textit{ISO} observations.
However, nothing concrete can be deduced by considering a single form of PAH. 
The observed value of D/H that varies from 3\% to 30\% \citep[]{Peeters04,Onaka14} indicates a mixture of deuterated PAHs, if
present in the ISM. In our recent study \citep[]{Mridu15}, a [D/H]$\rm_{sc}$
ratio has been derived for a set of different molecules
which is comparable to observation made by \textit{ISO} and \textit{AKARI}. The [D/H]$\rm_{sc}$ ratio is simply [D/H]$\rm_{int}$/[D/H]$\rm_{num}$
\footnote{[D/H]$\rm_{num}$=$\frac{\rm{no~of~D~atoms}}{\rm{no~of~H~atoms}}$}. In our present work, we have made similar calculations
for ionized forms of deuterated ovalene as ionization of molecules is likely to occur in the ISM. 
This has been compared with our previous results \citep[]{Mridu15}. Table~\ref{tab3} lists the [D/H]$\rm_{sc}$ 
ratios from our previous as well as present results. The molecules are chosen in terms of increasing size. If the observed
D/H ratio is lower than the calculated [D/H]$\rm_{sc}$ ratio, it suggests a mixture of pure, deuterated, deuteronated and/or
other substituted PAH molecules.
\begin{table*}[h]
 \centering
  \begin{minipage}{70mm}
 \caption{[D/H]$\rm_{sc}$ ratios calculated for DPAH$^+$ }
 \label{tab3}
 \begin{tabular}[c]{c|c}
 \hline \hline
 DPAH$^+$ & [D/H]$\rm_{sc}$ \footnote[7]{[D/H]$\rm_{sc}$=$\frac{\rm [D/H]_{int}}{\rm [D/H]_{num}}$} \\ \hline
Deuteronated pyrene \footnote[8]{\label{note1}\citep[]{Mridu15}}&  5.00 \\ \hline
Deuteronated perylene \footnoteref{note1} & 3.39 \\ \hline
Deuteronated coronene \footnoteref{note1} &  2.71  \\ \hline
DcorD$^+$ \footnoteref{note1}  & 1.43 \\ \hline
Deuterated ovalene cation\footnote[9]{\label{note2}present work} & 1.38 \\ \hline
Deuterated ovalene cation isomer 1\footnoteref{note2} & 1.63 \\ \hline
Deuterated ovalene cation isomer 2\footnoteref{note2} & 1.63 \\ \hline
Deuterated ovalene cation isomer 3\footnoteref{note2} & 1.50 \\ \hline
Deuteronated ovalene \footnoteref{note2}  &  0.57 \\ \hline
Deuteronated ovalene isomer 1\footnoteref{note2}  &  1.14 \\ \hline
Deuteronated ovalene isomer 2\footnoteref{note2}  &  2.43 \\ \hline
Deuteronated ovalene isomer 3\footnoteref{note2}  &  1.71 \\ \hline
DovaleneD$^+$ \footnoteref{note2}  &  0.93 \\ \hline
DovaleneD$^+$ isomer \footnoteref{note2}  &  1.47 \\ \hline
Deuteronated circumcoronene \footnoteref{note1} &  0.13 \\ \hline

\end{tabular}
\end{minipage}
 \end{table*}

D/H ratios provided by \citet[]{Peeters04} and \citet[]{Onaka14} may be
used to estimate the size of the deuterium-containing PAH variants in the ISM by comparing with the
\begin{figure*}[h]
\centering
\includegraphics[width=8cm, height=5cm]{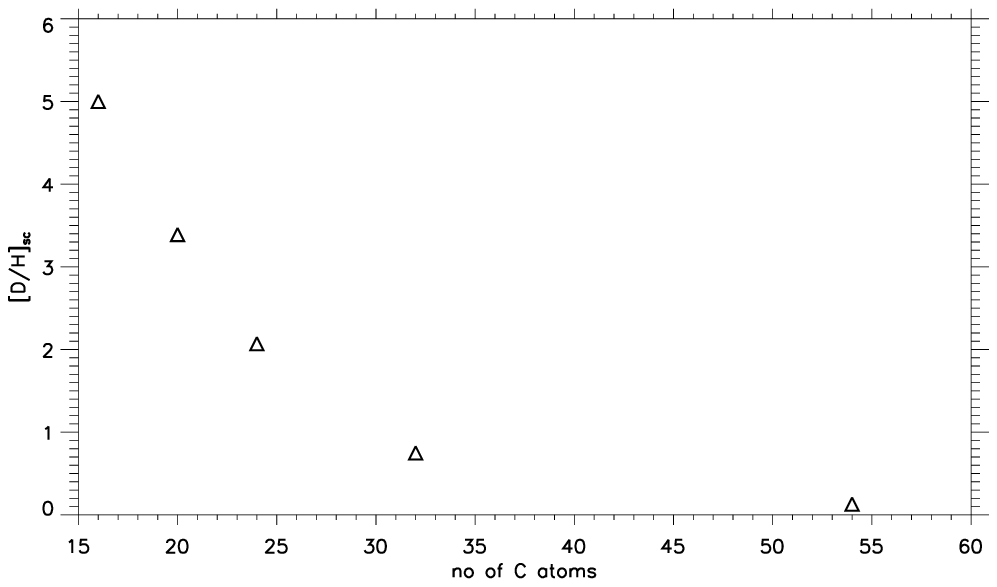}
\caption{[D/H]$\rm_{sc}$ ratio with increasing size, no of C atoms is directly proportional to size of PAH molecule}
\label{fig7}
\end{figure*}
theoretically obtained D/H ratio.
Our results from previous work \citep[]{Mridu15} and the present report
suggest a [D/H]$\rm_{sc}$ ratio (Table~\ref{tab3}) that is large compared to the \citet[]{Peeters04}
and \citet[]{Onaka14} observations except for deuteronated ovalene and 
deuteronated circumcoronene. The three isomers 
of deuteronated ovalene show a larger value of [D/H]$\rm_{sc}$. With increasing size, the [D/H]$\rm_{sc}$ ratio tends
to decrease (Figure~\ref{fig7}). For a particular molecule, addition of more than one deuterium may 
or may not lead to an increase in the [D/H]$\rm_{sc}$ ratio. [D/H]$\rm_{sc}$ ratios for deuteronated ovalene and 
deuteronated circumcoronene are in close proximity to the D/H ratio observed by \citet[]{Peeters04} 
and hence may be considered as feasible carriers for bands at $4-5~\mu \rm m$. 
[D/H]$\rm_{sc}$ ratios suggested by our work do not match with \textit{AKARI} observation. 
It is suggested that PAH molecules close to 100 carbon atoms may match the observed D/H ratio by \textit{AKARI}.

\citet[]{Doney15} observed H~\textsc{ii} regions in the Milky Way in near 
infrared in order to estimate the amount of deuterium in PAHs. They 
did not detect emission from deuterated PAHs towards all the sources and concluded 
that the deuteration of PAHs is not common. \citet[]{Doney15}
calculated emission spectrum for molecule with aliphatic group attached to the 
PAH structure and calculated
a D/H ratio in the range of 0.01$-$0.06 by taking the intensity ratio of 3.3 and 4.75 $\mu \rm m$ band.
Our approach is different from \citet[]{Doney15} and the [D/H]$\rm_{int}$ ratio is calculated
by taking the ratio of integrated band intensities  
of both aromatic and aliphatic $\rm C-D$ stretching to
integrated band intensities of both aromatic and aliphatic $\rm C-H$ stretching for our sample
molecules. The calculated [D/H]$\rm_{int}$ in our work for solo deuteration is in the range of 0.04-0.17. 
The molecules with an aliphatic side group however are feasible only in benign environments and may be destroyed
in a harsh interstellar environment such as fully evolved planetary nebulae or H~\textsc{ii} regions \citep[]{Bernstein96}.

\section{Conclusion}
Deuterium-containing PAH variants have been studied theoretically in relation to mid-infrared
emission bands. This report suggests PAH molecules with deuterium content as potential candidate carriers
for some of the observed UIR features in the ISM on basis of the band positions. 
PAHs with a D or D$^+$ give features in the $4-5~\mu \rm m$
region which arises purely due to the stretching of the $\rm{C-D}$ bond and hence may be considered
responsible for observed bands at $4.4~\mu \rm m$ and $4.65~\mu \rm m$. To gain further support, the [D/H]$\rm_{sc}$ ratio
for ovalene with D is estimated. On comparing this ratio from the present and previous reports with 
observations, it is realized that deuteronated ovalene and deuteronated circumcoronene agree 
with the observations of \citet[]{Peeters04}. \textit{AKARI} observations 
propose comparatively large PAHs (no. of C atoms~$\sim$~100) with low deuterium
content. PAHs with a suitable [D/H]$\rm_{sc}$ may be arranged accordingly for much more
expensive laboratory experiments which are of utmost
importance for assignment of carriers. The study of deuterium-containing PAHs is essential in order to measure D/H that 
will give insight into the history of star formation. This study can further be progressed to estimate  HD/H$_2$
ratio in interstellar gas. For reliable analysis, more experimental and observational evidence of 
interstellar deuterium is needed.
\section*{Acknowledgements}
We thank the anonymous referee for their critical comments that have helped
in improving the manuscript. MB is a junior research fellow in a SERB – DST FAST TRACK
project. AP acknowledges financial support from SERB DST FAST
TRACK grant (SERB/F/5143/2013–2014) and AP, TO and IS acknowledge
support from the DST – JSPS grant (DST/INT/JSPS/P-189/2014). 
AP thanks the Inter-University Centre for Astronomy
and Astrophysics, Pune for associateship. PJS thanks the Leverhulme Trust for award of 
a Research Fellowship and Leiden Observatory for hospitality that allowed completion of this work.
\section*{REFERENCES}
\def\aj{AJ}%
\def\actaa{Acta Astron.}%
\def\araa{ARA\&A}%
\def\apj{ApJ}%
\def\apjl{ApJ}%
\def\apjs{ApJS}%
\def\ao{Appl.~Opt.}%
\def\apss{Ap\&SS}%
\def\aap{A\&A}%
\def\aapr{A\&A~Rev.}%
\def\aaps{A\&AS}%
\def\azh{AZh}%
\def\baas{BAAS}%
\def\bac{Bull. astr. Inst. Czechosl.}%
\def\caa{Chinese Astron. Astrophys.}%
\def\cjaa{Chinese J. Astron. Astrophys.}%
\def\icarus{Icarus}%
\def\jcap{J. Cosmology Astropart. Phys.}%
\def\jrasc{JRASC}%
\def\mnras{MNRAS}%
\def\memras{MmRAS}%
\def\na{New A}%
\def\nar{New A Rev.}%
\def\pasa{PASA}%
\def\pra{Phys.~Rev.~A}%
\def\prb{Phys.~Rev.~B}%
\def\prc{Phys.~Rev.~C}%
\def\prd{Phys.~Rev.~D}%
\def\pre{Phys.~Rev.~E}%
\def\prl{Phys.~Rev.~Lett.}%
\def\pasp{PASP}%
\def\pasj{PASJ}%
\def\qjras{QJRAS}%
\def\rmxaa{Rev. Mexicana Astron. Astrofis.}%
\def\skytel{S\&T}%
\def\solphys{Sol.~Phys.}%
\def\sovast{Soviet~Ast.}%
\def\ssr{Space~Sci.~Rev.}%
\def\zap{ZAp}%
\def\nat{Nature}%
\def\iaucirc{IAU~Circ.}%
\def\aplett{Astrophys.~Lett.}%
\def\apspr{Astrophys.~Space~Phys.~Res.}%
\def\bain{Bull.~Astron.~Inst.~Netherlands}%
\def\fcp{Fund.~Cosmic~Phys.}%
\def\gca{Geochim.~Cosmochim.~Acta}%
\def\grl{Geophys.~Res.~Lett.}%
\def\jcp{J.~Chem.~Phys.}%
\def\jgr{J.~Geophys.~Res.}%
\def\jqsrt{J.~Quant.~Spec.~Radiat.~Transf.}%
\def\memsai{Mem.~Soc.~Astron.~Italiana}%
\def\nphysa{Nucl.~Phys.~A}%
\def\physrep{Phys.~Rep.}%
\def\physscr{Phys.~Scr}%
\def\planss{Planet.~Space~Sci.}%
\def\procspie{Proc.~SPIE}%
\let\astap=\aap
\let\apjlett=\apjl
\let\apjsupp=\apjs
\let\applopt=\ao
\bibliographystyle{elsarticle-harv} 
\bibliography{mridu}

\end{document}